\documentclass[conference,anonymous]{IEEEtran}
\IEEEoverridecommandlockouts
\usepackage{cite}
\usepackage{amsmath,amssymb,amsfonts}
\usepackage{algorithmic}
\usepackage{graphicx}
\usepackage{textcomp}
\usepackage{xcolor}
\usepackage{booktabs}

\def\BibTeX{{\rm B\kern-.05em{\sc i\kern-.025em b}\kern-.08em
    T\kern-.1667em\lower.7ex\hbox{E}\kern-.125emX}}
\begin{document}
\pagestyle{plain} 
\title{Intertwined Biases Across Social Media Spheres: Unpacking Correlations in Media Bias Dimensions}

\author{\IEEEauthorblockN{Yifan Liu}
\IEEEauthorblockA{\textit{School of Information Sciences} \\
\textit{University of Illinois} \\
\textit{Urbana-Champaign}\\
Champaign, U.S. \\
yifan40@illinois.edu}
\and
\IEEEauthorblockN{Yike Li}
\IEEEauthorblockA{\textit{School of Information Sciences} \\
\textit{University of Illinois} \\
\textit{Urbana-Champaign}\\
Champaign, U.S. \\
yikeli3@illinois.edu}
\and
\IEEEauthorblockN{Dong Wang}
\IEEEauthorblockA{\textit{School of Information Sciences} \\
\textit{University of Illinois} \\
\textit{Urbana-Champaign}\\
Champaign, U.S. \\
dwang24@illinois.edu}
}

\maketitle

\begin{abstract}
Media bias significantly shapes public perception by reinforcing stereotypes and exacerbating societal divisions. Prior research has often focused on isolated media bias dimensions such as \textit{political bias} or \textit{racial bias}, neglecting the complex interrelationships among various bias dimensions across different topic domains. Moreover, we observe that models trained on existing media bias benchmarks fail to generalize effectively on recent social media posts, particularly in certain bias identification tasks. This shortfall primarily arises because these benchmarks do not adequately reflect the rapidly evolving nature of social media content, which is characterized by shifting user behaviors and emerging trends. In response to these limitations, our research introduces a novel dataset collected from YouTube and Reddit over the past five years. Our dataset includes automated annotations for YouTube content across a broad spectrum of bias dimensions, such as gender, racial, and political biases, as well as hate speech, among others. It spans diverse domains including politics, sports, healthcare, education, and entertainment, reflecting the complex interplay of biases across different societal sectors. Through comprehensive statistical analysis, we identify significant differences in bias expression patterns and intra-domain bias correlations across these domains. By utilizing our understanding of the correlations among various bias dimensions, we lay the groundwork for creating advanced systems capable of detecting multiple biases simultaneously. Overall, our dataset advances the field of media bias identification, contributing to the development of tools that promote fairer media consumption. The comprehensive awareness of existing media bias fosters more ethical journalism, promotes cultural sensitivity, and supports a more informed and equitable public discourse.
\end{abstract}
\section{Introduction}
In our digital era with a proliferation of information across various social media platforms, media bias can significantly contribute to the reinforcement of stereotypes, discrimination and hatred. Beyond reinforcing existing social biases, media bias can resonate with and exacerbate inherent cognitive biases~\cite{Nickerson1998175}, leading to the formation of information bubbles~\cite{pariser2011filter}. The interaction between media and cognitive biases not only distorts public perception but also amplifies social divisions, which underscores the urgency of developing robust media bias identification systems to mitigate these effects. Despite being a critical issue, the definition of media bias has not reached a consensus for a long time, resulting in a range of different definitions~\cite{alessio2006, mullainathan}. A comprehensive literature review in recent research has sought to address this issue by proposing a unified definition of media bias. This definition includes a range of skewed portrayals that are systematically categorized into distinct media bias dimensions~\cite{spinde2024media}, highlighting the multi-faceted nature of media bias. In order to mitigate the negative effect of media bias, a range or frameworks are developed to assist automated media bias identification. 

The identification of media bias has primarily been addressed by the machine learning (ML) and natural language processing (NLP) communities~\cite{mediabias_detection}. In recent years, methods for identifying media bias have progressed from relying on hand-crafted features~\cite{recasens-etal-2013-linguistic, hand_crafted_wikipedia} to the employment of advanced transformer-based models~\cite{raza_dbias_2024, spinde_neural_MTL}. However, many existing research efforts still focus predominantly on detecting a \textit{single} type of media bias, typically evaluating models against benchmarks that assess only one dimension of bias~\cite{wang-2017-liar, Aksenov2021FinegrainedCO}. Such a narrow focus presents significant challenges for developing comprehensive bias identification systems: 1) there is a disproportionate focus within the media bias detection community on different bias dimensions, resulting in a lack of high-quality benchmark datasets along some bias dimensions~\cite{MBIB}, and 2) without a thorough understanding of the various media bias dimensions, it is challenging to develop a bias identification system that is capable of jointly detecting and analyzing multiple bias dimensions. 

In response to these problems, several recent efforts have been made to address the multi-dimensional nature of media bias~\cite{MBIB, spinde2024media, spinde_neural_MTL}. Beyond the traditional focus on single-dimensional media bias, recent work has expanded the discourse by developing a comprehensive taxonomy of media bias dimensions, which unifies the definition of multi-dimensional media bias~\cite{spinde2024media}. Building on this definition of media bias, Wessel et al. surveyed and organized a series of benchmark datasets for each bias dimension including linguistic bias~\cite{LinguisticBias}, political bias~\cite{feldman_political_2013}, gender bias~\cite{MBIB}, etc. While these efforts inspire the creation of more comprehensive bias identification systems, we note that there is yet a comprehensive investigation to improve the understanding of inter-correlations among different media bias dimensions. 

To this end, we create a media bias benchmark where multiple dimensions of media biases are annotated and analyzed across different topic domains. Building on the bias specification of prior work~\cite{MBIB}, we select a subset of bias dimensions that are better tailored for collected social media data. We report our bias specification in Section~\ref{subsec:biasDim}, where we detail the definitions of bias dimensions involved in our discussion. In our dataset creation process, we collect social media posts over the past 5 years from YouTube and Reddit. With rigorous statistical analysis, we provide insights by answering the following research questions: 1). How do the prevalence of different media bias dimensions vary across various domains? 2). How do different media bias dimensions correlate with each other within each topic domains? 3). How do correlations among different media bias dimensions vary across various topic domains?

Overall, we find politics domain differs from all other topic domains with significantly higher proportions of biased contents. Furthermore, we observe domain-specific patterns in the stylistic expressions of bias within posts. For instance, in the politics domain, biased posts often manifest through manipulating the narrative to serve a particular agenda, whereas in the sports domain, biased many manifest through the use of specific word choices or jargon that convey prejudices. Furthermore, our analysis reveals that when aggregating data along the temporal dimension, the correlations between different bias dimensions are not static but dynamically fluctuate in response to spikes in discussion volume. Our main contributions are:
\begin{itemize}
    \item We have created a cross-platform (YouTube and Reddit) corpus spanning five distinct topic domains.
    \item We annotated YouTube comments to create a multi-dimensional media bias identification benchmark dataset.
    \item We conducted a thorough analysis of the correlations and domain-specific discrepancies within our annotated datasets, significantly enhancing our understanding of the interactions between various media bias dimensions.
    \item Our work is the first dataset to account for the joint occurrence of multiple bias dimensions. 
\end{itemize}


\section{Related works}
\label{sec:relatedWorks}
\subsection{Media Bias}
Media bias is a broad concept that has been investigated since the 1950s~\cite{earlyMediaBias}. It has been extensively studied over the years under various definitions. For instance, Spinde et al. define media bias as slanted news coverage or internal bias, encompassing both report-level and content-level biases~\cite{SPINDE2021102505}. In a more recent research, media bias is also characterized as the prejudiced portrayal in communication~\cite{communicationBias}, focusing primarily on content. For more fine-grained categorizations, media bias can be divided into gate-keeping bias, coverage bias, and statement bias~\cite{alessio2006}; it can also be split into ideology bias and spin bias~\cite{mullainathan}. The variety of definitions and categories in prior studies on media bias have posed challenges for the development of media bias analysis and biased content detection. To address this, Spinde et al. conducted a comprehensive literature review across multiple research fields of media bias, providing a thorough taxonomy~\cite{spinde2024media}. In subsequent work that aimed at advancing automated bias detection schemes, Wessel et al. surveyed over 115 datasets and established a Media Bias Identification Benchmark(MBIB) related to various dimensions of media bias as summarized by prior work~\cite{spinde2024media}. Our research builds upon the existing taxonomy in MBIB, with a focus on social media platforms.

\subsection{Media Bias Dimensions}
Compared to previous studies on media bias, our work emphasizes the multi-dimensional aspects of media bias and the domain-specific occurrences of bias dimensions on social media platforms. Recent research has utilized the interrelationships between different types of biases as a foundation for developing more robust bias identification systems~\cite{MBIB}. Specifically, this approach is applied within the multi-task learning (MTL) framework~\cite{chen2021multitask, horych2024magpie}, which provides a joint optimization framework for various media bias dimensions. It is important to note that both task selection and data selection play critical roles in the success of MTL~\cite{bingel-sogaard-2017-identifying, ruder-plank-2017-learning}. To effectively address such challenges, automated task-selection algorithms are considered to be a promising enhancement~\cite{ma-etal-2021-gradts}. However, gradient-based automated task selection schemes, which rely on monitoring the training dynamics of different sub-tasks, are susceptible to discrepancies between data sources and could require a large number auxiliary tasks to perform effectively~\cite{horych2024magpie}.

\subsection{Comparison with Other Media Bias Datasets}
\label{sec:comparison}
Within the field of media bias research, various datasets are specifically designed to support the analysis of a single media bias type, employing multi-level labeling to capture its nuances. For example, RTGender dataset adopt a multi-categorical labeling to characterize different aspects of gender bias~\cite{voigt-etal-2018-rtgender}. Similarly, prior research on political bias categorizes texts into five distinct political tenancies, ranging from \textit{left} to \textit{right}~\cite{Aksenov2021FinegrainedCO}. For more nuanced labeling, CMSB employs a continuous sexism scale to measure subtleties in Twitter data\cite{sexim}. While such fine-grained categorization in each bias dimension provides a detailed view of media bias, our work primarily focuses on exploring inter-correlations among different bias dimensions.

Similar to our work, multidimensional bias dataset collects dataset with a bias dimension specification based on hidden assumptions, subjectivity and representation tendencies~\cite{multidimension}. However, multidimensional bias dataset has a focus on news articles with a special emphasis on political tendencies, while our work focuses on the social media space with a wider range of topics. Recent media bias identification benchmark(MBIB) summarizes a rich list of publicly available datasets following a set of media bias dimension specification~\cite{MBIB, spinde2024media}. Despite the well rounded bias dimensions discussed, MBIB provides a single label for each post, limiting the analysis of across-dimension analysis of media bias.

Compared to other existing bias identification datasets, our dataset is the first to account for the joint occurrence of multiple bias dimensions with a joint labeling scheme. Additionally, we have intentionally segregated data collections across different domains using general keyword choices. In this section, we discuss the distinctions and advantages of our dataset compared to the most related previous works. 

\section{Dataset Creation}
In this section, we elaborate on our dataset creation process, which includes the following steps: 1) Retrieval of domain-specific data; 2) Examination of bias dimensions leading to the development of our bias specification framework; and 3) Generation of multi-dimensional bias labels. Furthermore, we compare our dataset with existing datasets to underscore the research gap in the study of intertwined media biases. Our data collection process is designed to investigate dimensions of media bias within and across various domains, namely politics, healthcare, sports, entertainment and job \& education. 

\subsection{Data Retrieval and Filtering}
To conduct a comprehensive analysis of media bias across social media platforms, we gathered comments and posts from specific domains on Reddit and YouTube. To ensure the effective representation of each domain, we used Google Trends to guide the selection of representative keywords for each topic domain. For instance, "COVID" has emerged as the most frequently searched term in the healthcare sector over the past five years, indicating its significant public exposure and suitability as a keyword. Our keywords used is illustrated in table~\ref{table:keywords}. We utilize the keyword search functionality of Reddit's and YouTube's data collection APIs to gather pertinent social media posts spanning the last five years. For Reddit, we amassed over 30,000 posts per domain. For YouTube, we collected approximately 2,000 comments per domain. It is important to note that our analysis on YouTube comments only includes comments directly related to the video topics, while all sub-comments replying to other comments have been excluded. In this way, we ensure that our analysis retains most domain-specific comments, while eliminating any outliers resulting from off-topic discussions. Building on top of our collected data, we prepare our raw data for media bias annotations by filtering out non-English posts and posts that are too long ($>200$ words). While prior work commonly remove emojis, we convert emojis to text tokens to keep the semantics of posts intact in bias annotations, which we provide more details in section~\ref{sec:annotation}.

\begin{table*}[htbp]
\caption{Data Collection by Domain}
\label{table:keywords}
\centering
\begin{tabular}{llll}
\toprule
Domain & Keywords & YouTube Comments & Reddit Posts\\
\midrule
Politics & election contest, election result, voting & 1993 & 12842 \\
Healthcare & COVID, pain injury, symptom & 2580 & 14753 \\
Sports & NBA, NFL, MLB & 2398 & 24728 \\
Entertainment & film, lyrics, episodes & 998 & 40852 \\
Job \& Education & career, college, job & 1455 & 36966 \\
\bottomrule
\end{tabular}
\end{table*}

\subsection{Bias Dimension Specifications}
\label{subsec:biasDim}
In our analysis, we ground our investigation under the umbrella of media bias introduced in recent works~\cite{MBIB, spinde2024media}. In our exploration, we investigate a subset of media bias types summarized in prior work~\cite{MBIB}. Specifically, we choose the bias types that are defined at post-level with an emphasis on lexical features that can be efficiently captured by transformer-based models for more accurate analysis. We review the bias dimensions under investigation and provide their respective definitions below.

\subsubsection{Linguistic Bias~\cite{LinguisticBias}} is characterized by biases stemming from lexical features, manifesting through the specific choices of words that reflect social-category cognition attributed to described groups or individuals. 

\subsubsection{Political Bias~\cite{feldman_political_2013}} refers to the expression of a particular political ideology, altering the political discourse by emphasizing certain viewpoints over others, thereby shaping the agenda of political discussions and debates \cite{iyengar_media_2009}.

\subsubsection{Gender Bias~\cite{MBIB}} encompasses discrimination against gender groups as manifested in textual content. This form of discrimination can appear through underrepresentation, reliance on stereotypes, or negative portrayals—each of which contributes to the perpetuation of gender inequality.

\subsubsection{Hate Speech~\cite{DavidsonWMW17, hatexplain}} is characterized by language that expresses hatred towards a group or individual(s). On social media, hate speech represents not only a form of biased text but also a significant cultural threat that has been linked to crimes against minorities~\cite{hateImpact}.

\subsubsection{Racial Bias} is defined as texts that express negative or positive descriptions towards racial groups, which could play a pivotal role in shaping public discourse and policy-making related to race relations and diversity~\cite{dixon_priming_2007}.

\subsubsection{Text-level Context Bias~\cite{MBIB}} is characterized by the strategic use of specific words and statements within texts to craft a narrative that favors a particular viewpoint of an event. This includes presenting a skewed description of incidents, thereby biasing the narrative towards one side of an argument. 

To examine correlations across different dimensions of media bias, we selected a subset of biases as identified in prior research~\cite{MBIB}. Our focus is on dimensions of media bias that are defined solely by the content of social media posts, independent of broader contextual factors. Consequently, we exclude the following from our analysis: 1) Fake news, as its detection often requires both lexical analysis and knowledge-intensive classification, relying on auxiliary information such as user profiles; 2) Cognitive bias, which is predicated on the reader's pre-existing beliefs and is difficult to ascertain from text content alone; and 3) Reporting-level context bias, which pertains to how social platforms may selectively present events and content, falling outside our investigative scope. For comprehensive definitions of these excluded media bias dimensions, we direct readers to related works~\cite{MBIB, spinde2024media}.

\subsection{Bias Annotation}

To annotate the social media posts we collected, we employed a mixed approach combining manual and automated annotations. For each domain dataset, 100 samples were annotated by two annotators. Each annotator was responsible for annotating 60 samples, with 20 samples overlapping between the two annotators for consistency checks. In table~\ref{table:kappa}, we report the size of filtered dataset, together with the inter-rater agreement score (Cohen's $\kappa$) for each domain. Across all bias dimensions, we observe substantial inter-rater agreement (Cohen's $\kappa > 0.8$), with the conflicting annotations being reviewed and inspected. Building on the manual annotations, we evaluated two groups of automated annotations: 1) a shallow pretrained models trained on existing benchmark bias evaluation datasets; and 2) zero/few-shot annotations from pretrained large language model~\cite{ouyang2022training}. Our results indicate that while smaller-scale transformer baselines achieve good robustness and generalizability with cross-validated train-test splits, they exhibit poor generalization when applied to the noisy social media posts we collected along some bias dimensions. Overall, we utilize the predictions generated by the best performing considering both shallow models and LLMs evaluated on our manually annotated test set in our further analysis of bias dimensions. The details of our pre-processing and annotation process and experiments are reported in Section~\ref{experiments}.
\begin{table}[htbp]
\caption{Inter-Rater Agreement for Bias Dimensions}
\label{table:kappa}
\label{table_example}
\centering
\begin{tabular}{l|l}
\hline
\textbf{Category} & \textbf{Cohen's $\kappa$ Score} \\
\hline
Gender Bias & 1.00 \\
Racial Bias & 1.00 \\
Hate Speech & 0.82 \\
Linguistic Bias & 0.93 \\
Text-Level Context Bias & 0.89 \\
Political Bias & 0.81 \\
\hline
\end{tabular}
\end{table}
\label{sec:annotation}

\begin{table*}[ht]
\caption{Weighted F1-Scores Across Different Evaluation Domains}
\centering
\begin{tabular}{l|ccccc|c}
\toprule
\textbf{Bias Dimension} & \textbf{Politics} & \textbf{Sports} & \textbf{Healthcare} & \textbf{Job\&Education} & \textbf{Entertainment} & \textbf{Model} \\
\midrule
Gender Bias & 0.90 & 0.99 & 0.99 & 0.97 & 0.88 & GPT-Turbo-3.5\\
Racial Bias & 0.97 & 0.98 & 1.00 & 0.93 & 0.96 & GPT-Turbo-3.5\\
Hate Speech & 0.77 & 0.85 & 0.92 & 0.95 & 0.91 & Roberta-Twitter\\
Linguistic Bias & 0.63 & 0.75 & 0.81 & 0.84 & 0.63 & GPT-Turbo-3.5\\
Text-level Context Bias & 0.56 & 0.77 & 0.86 & 0.84 & 0.87 & ConvBert \\
Political Bias & 0.64 & 1.00 & 0.94 & 0.98 & 1.00 & GPT-Turbo-3.5\\
\bottomrule
\end{tabular}
\label{tab:f1_scores}
\end{table*}

\begin{table*}[ht]
\caption{Number of Biased Content Along Different Dimensions from Automated Annotation \\with Percentage of Total Posts Shown in Brackets (\%)}
\centering
\begin{tabular}{c|c|c|c|c|c}
\toprule
\textbf{Bias Dimension} & \textbf{Politics} & \textbf{Sports} & \textbf{Healthcare} & \textbf{Job\&Education} & \textbf{Entertainment} \\
\midrule
Gender Bias & 91 (4.61) & 76 (3.21) & 77 (3.09) & 72 (5.15) & 44 (4.53) \\
Racial Bias & 99 (5.02) & 39 (1.64) & 30 (1.21) & 35 (2.50) & 14 (1.44) \\
Hate Speech & 464 (23.54) & 330 (13.91) & 244 (9.82) & 172 (12.29) & 101 (10.40) \\
Linguistic Bias & 499 (25.32) & 501 (21.13) & 382 (15.37) & 239 (17.08) & 192 (19.77) \\
Text-level Context Bias & 480 (24.35) & 204 (8.60) & 247 (9.94) & 123 (8.79) & 56 (5.77) \\
Political Bias & 578 (29.32) & 35 (1.48) & 60 (2.42) & 18 (1.29) & 3(0.31) \\
\midrule
Total Posts & 1971 & 2371 & 2484 & 1399 & 971\\
\bottomrule
\end{tabular}
\label{tab:label_count}
\end{table*}
\section{Experiments}
\label{experiments}
\subsection{Automated Annotation}
For automated annotation, we evaluate the empirical performances of a collection of shallow pretrined models (ConvBERT, GPT2, Bart, RoBERTa-Twitter and ELECTRA) and existing LLMs (GPT-3.5-turbo) using our manually annotated data as test dataset. We observe that the bias identification task presents as an imbalanced classification challenge, where less than 20\% of the annotations across all bias dimensions are positive. To account for class imbalance, we use weighted average F1 score as our evaluation metrics for all our automated annotations following practices in prior works~\cite{harbecke2022microf1, MBIB}. 

For the shallow pretrained models, we select the best baseline models validated on MBIB~\cite{MBIB} datasets per bias identification tasks. Our training pipeline and pretrained models are implemented with PyTorch~\cite{torch} and HuggingFace Transformers~\cite{huggingface}. For the models trained on MBIB datasets, we adopt a random 5-fold train-validation split with a 75\% data used for training and 25\% data used for validation, where the best-performing model in validation set is used for evaluation on our collected social media posts. 

The prediction performances of the final models used for automatic labeling are reported in Table~\ref{tab:f1_scores}. Notably, in the politics domain, the identifications of \textit{linguistic bias}, \textit{text-level context bias}, and \textit{political bias} shows lower prediction performance due to the nuanced definitions of these biases. This challenge is also reflected in our inter-rater agreements as shown in Table~\ref{table:kappa}, which suggest that these bias dimensions are inherently more difficult to predict. Nevertheless, we observed improved performance in non-politics domains, indicating the model's efficacy of our annotation scheme. Overall, we conclude that our trained models are capable of effectively detecting the bias in the targeted dimensions when evaluated on our manually annotated test sets. 

In our evaluation, shallow pretrained models show strong performance in detecting \textit{hate speech} and \textit{text-level context bias}. However, performance significantly declines for other bias dimensions when applying models trained on MBIB datasets to our social media dataset. For example, the best racial bias detector trained on MBIB datasets achieves an average weighted F1 score of 0.82, but drops to 0.12 on our dataset, indicating a failure to generate meaningful predictions. This performance drop, ranging from 0.1 to 0.69 for biases other than \textit{hate speech} and \textit{text-level context bias}, suggests that existing models struggle with the complexity and variability of our collected data. The robust detection of \textit{hate speech} likely benefits from focused prior research with high-quality datasets. The observed variances in model effectiveness across different biases highlight the need for a deeper understanding of each bias dimension, underscoring the importance of our study in advancing bias identification technologies.

On the other hand, we observe consistent performance of bias identification across all bias dimensions from the zero-shot predictions of the large language model (GPT-3.5-turbo). In our prompting scheme, we tested in-context learning with (0, 1, 3) shots, where similar prediction performances were noted across these zero/few-shot settings. We have chosen to use 1-shot in-context learning for our prediction generation pipeline. Across all types of bias identification, the LLM excels at predictions that focus on lexical features or specific word choices such as \textit{linguistic bias}, but it struggles with predicting more complex, semantically rich bias dimensions, such as \textit{text-level context bias}. Overall, we selected the models that performed best on our manually annotated test sets to serve as our automated annotators. Details of our model selections are reported in Table~\ref{tab:f1_scores}.

\subsection{Distribution Shift}
\begin{figure}[htb!]
\centerline{\includegraphics[width=0.4\textwidth]{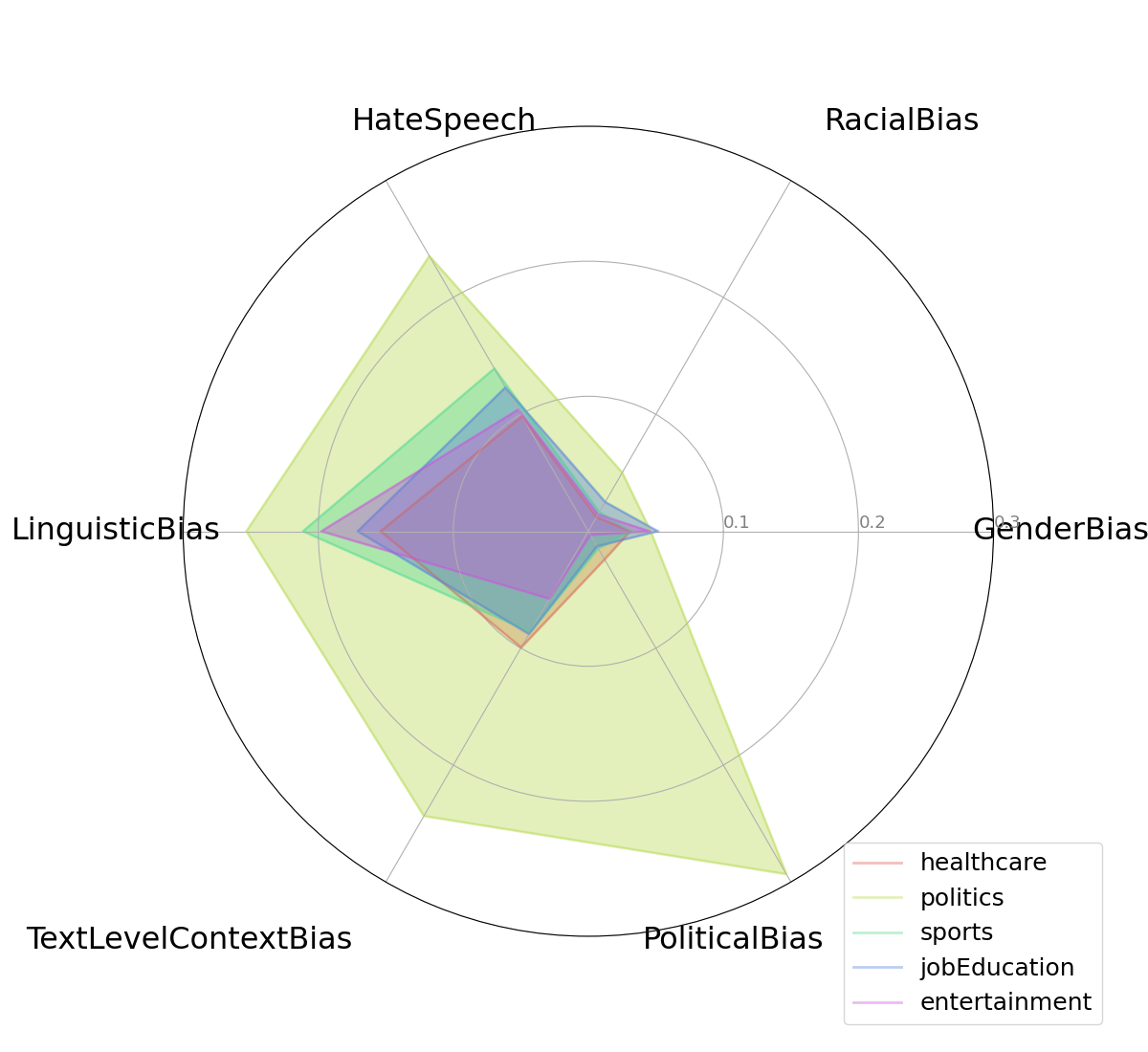}}
\caption{Radar plot illustrating the distribution of biased content across various dimensions. Notably, the political domain exhibits significantly higher proportions of various types of bias compared to other domains.}
\label{fig:radar}
\end{figure}

For all our annotated data, we report the number of samples and the percentages of their occurrences across different bias dimensions in Table~\ref{tab:label_count}. As illustrated in Figure~\ref{fig:radar}, while most domains exhibit similar bias elements across various dimensions, the politics domain is characterized by a significantly higher proportion of biased content, particularly notable in \textit{text-level context bias} and \textit{political bias}. Specifically, as shown in the radar plot (Figure~\ref{fig:radar}), the area representing the bias proportion in the politics domain nearly encompasses the areas of all other domains. In addition, we observe that in most domains, the shapes of radar plot indicate that biased content primarily manifests as linguistic biases, characterized by the use of specific discriminatory words. However, in the politics domain, biased content emerges through both specific word choices (\textit{linguistic bias}) and skewed descriptions (\textit{text-level context bias}) with similar proportions as shown in Table~\ref{tab:label_count}. 

\subsection{Correlated Bias Dimensions}
\begin{figure*}[htb!]
\centerline{\includegraphics[width=\textwidth]{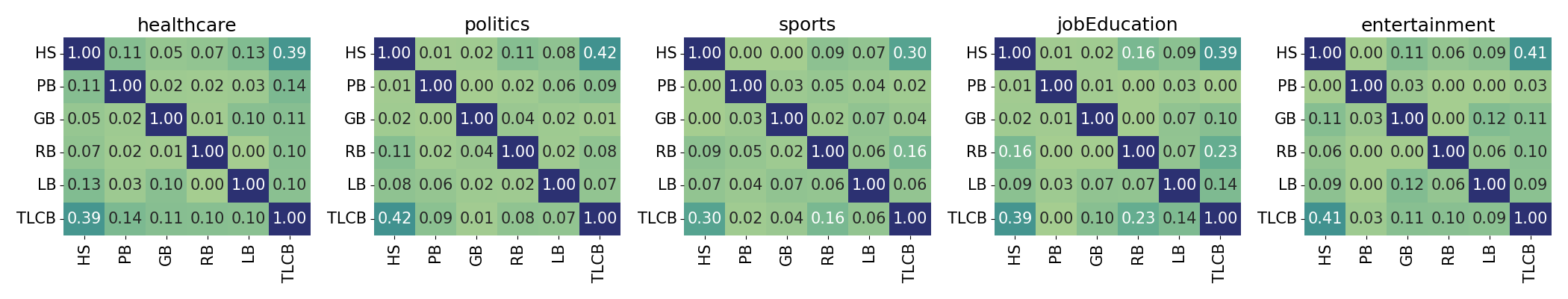}}
\caption{Correlation heatmap for each bias dimension of different domains, calculated using Cramér's $\mathcal{V}$. Higher values indicate stronger correlations. Abbreviations: HS for Hate Speech, PB for Political Bias, GB for Gender Bias, RB for Racial Bias, LB for Linguistic Bias and TLCB for Text-level Context Bias.}
\label{fig:heatmap_raw}
\end{figure*}
Based on the definitions of our media bias dimensions, we categorize them into two groups: i) style-based bias (\textit{linguistic bias, text-level context bias}), which indicates how biased texts are phrased, and ii) content-based bias (\textit{hate speech, gender bias, political bias, racial bias}), which refers to the different aspects or subjects of bias observable within the content. We note that while both groups of bias contribute to the spectrum of media bias, they often lead to different interpretations and thus may inspire the development of distinct bias identification frameworks. Consequently, in the subsequent discussion, we analyze and interpret the correlations between bias dimensions both within and between these two groups. For all collected posts, we utilize the binary labels generated for each bias dimension to analyze their co-occurrence. We begin by exploring the correlation between different bias dimensions within a topic domain. Specifically, we conduct pairwise chi-square tests to examine these relationships and report the Cramer's $\mathcal{V}$ value for each bias dimension pair as shown in figure~\ref{fig:heatmap_raw}. For all pairwise correlations in the subsequent discussion, the associated Chi-squared independence tests show strong statistical significance, with highest $p$-value being $1.87 \times 10^{-3}$. 

In order  to understand how media bias are expressed in different domains, we first investigate the correlations we observe associated with style-based bias dimensions. Specifically, we observe that for all topic domains, \textit{text-level context bias} is mostly associated with \textit{hate speech}. In sports and job \& education domain, we observe \textit{hate speech} also has a moderate positive correlation with racial bias with Cramer's $\mathcal{V} > 0.15$. Unlike \textit{text-level context bias}, \textit{linguistic bias}, which primarily focuses on specific word choices, does not exhibit a clear positive correlation with some specific types of biases, but more evenly correlated with all content-based bias dimensions. This observation suggests that content-based bias dimensions, particularly hate speech, are more often expressed through biased descriptions rather than specific biased terms. 

We further extend our discussion to correlations within the collection of content-based biases. In the subsequent discussion, we focus exclusively on correlations that are relatively strong, specifically where Cramer's $\mathcal{V} > 0.1$. Among the content-based bias dimensions, unlike other content-based biases, we observe \textit{hate speech} often coexists with other types of content-based biases. Across all topic domains, we observe that \textit{hate speech} is most significantly correlated with \textit{political bias} in healthcare, \textit{racial bias} in sports, \textit{racial bias} in politics, \textit{racial bias} in job \& education, and \textit{gender bias} in entertainment. Interestingly, apart from the (\textit{hate speech}, \textit{racial bias}) correlation in sports domain which is less significant than others, the aforementioned correlations may reflect the nuanced ways in which content creators and social media users engage with topics sensitive to identity and political context. For instance, in the healthcare domain, political discussions often intersect with deeply polarized issues such as healthcare policy and reproductive rights, which often incite hate speech. In politics and job \& education, racial discussions usually evoke strong biases, potentially escalating into hate speech. These correlations suggest that hate speech often serves as a marker for contentious social and political issues that stir significant public emotion and debate.

\subsection{Time Series Analysis}
\begin{figure*}[htb!]
\centerline{\includegraphics[width=1.1\textwidth]{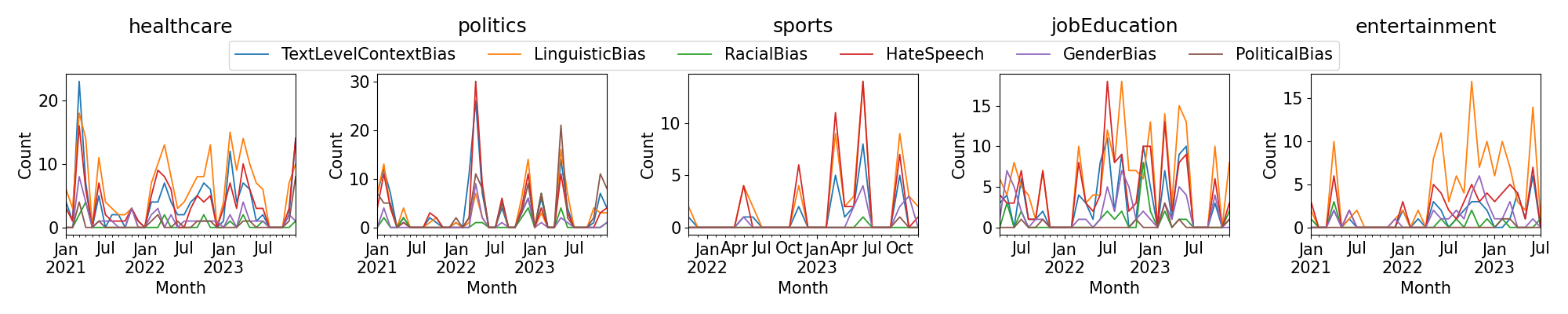}}
\caption{Line plot visualizations of monthly aggregated counts of bias dimensions. Key observations include: 1) Hate speech manifests in varying proportions of the two types of style-based bias dimensions across different domains. 2) Notable surges in aggressive biases are observed in specific months within the politics domain, supporting our hypothesis that biases in these domains are more event-driven compared to others.}
\label{fig:lineplots}
\end{figure*}
Beyond holistic analysis, social media is a rapidly evolving field characterized by swift changes in content and interaction patterns. In this context, we underscore the importance of incorporating temporal analysis to capture these dynamic changes over time. We aggregate the amount of biased contents within each bias dimension at one-month intervals. We then analyze the time series of these bias dimensions across different topic domains as well as the correlations between aggregated bias dimensions within the same domain. Our overall aggregated data is presented in Figure~\ref{fig:lineplots}. Beyond general line plot interpretation, we conduct statistical tests to further investigate differences across domains. For every comparison pair, we standardize the time series data to a common time frame. Subsequently, we perform domain-pairwise comparisons by conducting t-tests across all bias dimensions. For intra-domain bias analysis, we report insights from the top-2 correlated pairs, measured using Pearson’s correlation.

We first investigate the strong (style, content)-based correlation pairs. Most notably, diverging from the typically dominant correlation of (\textit{hate speech}, \textit{text-level content bias}), \textit{linguistic bias} emerges as the most closely correlated style-based bias with \textit{hate speech} with a Pearson's coefficient of $0.86$ in the aggregated time series for the entertainment domain. Furthermore, the month-interval aggregation significantly strengthens the (\textit{hate speech}, \textit{linguistic bias}) correlation across all domains, with the lowest Pearson's coefficient observed being $0.77$ in the politics domain. The observation that temporal aggregation enhances the visibility of (\textit{hate speech}, \textit{linguistic bias}) correlation is likely due to the smoothing of outliers and noise in the data. The observed correlation in the entertainment domain, in particular, might reflect a trend where hate speech is more frequently expressed through nuanced language choices rather than content cues.

For content-based bias dimensions, we observe that aggregation introduces distinct correlations between bias dimensions. For clarity, we refer to correlations observed in monthly aggregated bias counts as 'short-term correlations,' and those across the entire dataset as 'long-term correlations.' We summarize the differences in the highest correlated pairs of biases as follows: 1) In the politics domain, the most closely correlated pair is (\textit{hate speech, gender bias}) with a Pearson's coefficient of $0.92$, which is significantly higher than the correlation between (\textit{hate speech, political bias}) with a coefficient of $0.69$. 2) In the sports domain, \textit{gender bias} and \textit{racial bias} show the closest correlation. 3) In job \& education domain, \textit{hate speech} is closely correlated with both \textit{gender bias} and \textit{racial bias} with Pearson's coefficients of $0.54$ and $0.53$ respectively. 

Overall, temporal aggregation enhances the sensitivity of our analysis to surges in social media posts and their associated biased content by focusing on temporally localized relationships. The aggregation's sensitivity to fluctuations may reveal discrepancies between the generally observed correlations of bias dimensions and those correlations that are particularly reactive to sudden increases in biased content. Therefore, variations in the top correlated pairs of bias dimensions could highlight how transient social media trends can affect the perceived interrelationships of bias dimensions. Through qualitative examination, we found that surges in biased social media posts often relate to specific social events. For example, during the congressional district elections in April and May 2022, there was a noticeable increase in occurrences of \textit{gender bias}, \textit{political bias}, and \textit{hate speech}, indicating a connection between these events and increased bias in social media posts.


\subsection{Implications to Multi-dimensional Bias Identification}
We summarize our findings and discuss their implications for the development of deep learning-based multi-dimensional bias identification systems as following: We have observed that \textit{hate speech} consistently shows strong correlations with certain domain-specific content-based bias types across bias identification tasks. Furthermore, our time series analysis indicates that short-term bias dimensions may exhibit different correlations from those observed in a longer-term context.

Building on these observations, we propose that adaptive multi-task learning targeted at highly correlated bias dimensions, could improve biased content detection. One approach could dynamically select combinations of tasks that maximize the sharing of mutual knowledge, thereby enhancing performance across related tasks. Additionally, we argue that effectively detecting and leveraging short-term surges in biased content can provide valuable priors for bias detection systems. These surges often reflect transient societal events or changes in public discourse, suggesting that a dynamic modeling approach could be crucial. By adapting to these fluctuations assisted by temporal analysis and event-level modeling, bias detection systems could become more responsive and accurate in real-world scenarios. We envision such multi-task learning frameworks could not only refine the detection mechanisms but also contribute to more nuanced understandings of bias dynamics over time, thereby supporting more informed and effective interventions against bias in social media.

\section{Discussion}
For the verification of our automated annotation scheme, we primarily rely on the alignment of model predictions to the manual annotations provided by two annotators. Notably, identifying biased content in social media can be inherently challenging, even for human annotators. In some cases, media bias benchmark datasets involve expert annotators or those with specialized training~\cite{wang-2017-liar, vidgen-etal-2021-introducing}. Despite these challenges, our dataset offers a unique variety compared to existing media bias benchmarks. Future research could enhance annotation with the assistance of experts in linguistics and social sciences. Additionally, employing crowdsourcing platforms may provide a viable source for further annotations. However, it is worth noting that a big concern of crowdsourcing approach is the data quality~\cite{MBIB}. Future work could take a hybrid expert-crowd approach to achieve reliable and scalable annotation.

Furthermore, while our current study includes a collection of social media posts from various domains, our annotations and analyses are confined to YouTube comments in order to maximize the domain specificity. This focus may limit our understanding of how biases manifest across different platforms. Future research could potentially explore cross-platform differences in the expression of intertwined bias dimensions. Such an investigation would be critical for developing a more comprehensive understanding of media biases, as each platform has unique user demographics, interaction modes, and content dissemination mechanisms, which could influence the nature and extent of biases. For example, X's character limit and real-time communication style might emphasize different biases compared to YouTube.


\section{Conclusion}
Our study aims to advance the understanding of media bias by introducing and investigating a social media dataset that spans multiple domains and bias dimensions, collected from YouTube and Reddit over the past five years. Moreover, our findings reveal significant differences in how biases are expressed across various domains such as politics, sports, and healthcare. We also discover fluctuations in the correlations between bias dimensions in response to surges in social media posts. These findings underscore the complex and evolving nature of media bias and lay the foundation for the future development of multi-dimensional bias identification systems. By advancing investigations into media bias, we hope to equip both researchers and practitioners with the tools necessary to address and mitigate the impacts of media bias, ultimately fostering a fairer media environment.

\bibliographystyle{ieeetr}
\bibliography{ref.bib}
\vspace{12pt}
\end{document}